\documentstyle[11pt,aaspp4,psfig]{article}
\tighten
\def\kms{km~s$^{-1}$~}
\def\etal{et al.~}
\def\HI{\ion{H}{1}~}
\def\la{\mathrel{\hbox{\rlap{\hbox{\lower4pt\hbox{$\sim$}}}\hbox{$<$}}}}
\def\ga{\mathrel{\hbox{\rlap{\hbox{\lower4pt\hbox{$\sim$}}}\hbox{$>$}}}}

\begin{document}

\title{\ion{H}{1} Bright Galaxies in the Southern Zone of Avoidance} 

\author{
P.A.\ Henning,\altaffilmark{1}
L.\ Staveley-Smith,\altaffilmark{2}
R.D.\ Ekers,\altaffilmark{2}
A.J.\ Green, \altaffilmark{3}
R.F.\ Haynes, \altaffilmark{2}
S.\ Juraszek,\altaffilmark{3}
M.J.\ Kesteven,\altaffilmark{2}
B.\ Koribalski,\altaffilmark{2}
R.C.\ Kraan-Korteweg,\altaffilmark{4}
R.M.\ Price,\altaffilmark{1}
E.M.\ Sadler, \altaffilmark{3}
A.\ Schr\"{o}der,\altaffilmark{5}
}

\altaffiltext{1}{Institute for Astrophysics, University of New Mexico,
800 Yale Blvd, NE, Albuquerque, NM 87131, USA}
\altaffiltext{2}{Australia Telescope National Facility, CSIRO, P.O.\ 
Box 76, Epping, NSW 2121, Australia}
\altaffiltext{3}{School of Physics, University of Sydney, NSW 2006,
Australia}
\altaffiltext{4}{Departamento de Astronomia, Universidad de Guanajuato,
Apartado Postal 144, Guanajuato, Gto 36000, Mexico}
\altaffiltext{5}{Observatoire de la C\^{o}te d'Azur, B.P. 4229,
06304 Nice Cedex 04, France}

\begin{abstract}

A blind survey for \ion{H}{1} bright galaxies in the southern
Zone of Avoidance, ($212^\circ \le \ell \le 36^\circ$,
$|b|$~$\leq$~5$^\circ$), has been made with the 21~cm multibeam
receiver on the Parkes 64~m radiotelescope.
The survey, sensitive to normal spiral galaxies to a distance
of $\sim$ 40 Mpc and more nearby dwarfs, detected 110 galaxies.
Of these, 67 have no counterparts cataloged in the NASA/IPAC
Extragalactic Database.  
In general, the uncataloged galaxies lie behind thicker obscuration than
do the cataloged objects.
All of the newly-discovered galaxies have \ion{H}{1} flux integrals
more than an order of magnitude lower than the Circinus galaxy.
The survey recovers the Puppis cluster and foreground group (Kraan-Korteweg
\& Huchtmeier 1992), and the Local Void remains empty.
The \HI mass function derived for the sample is satisfactorily fit
by a Schechter function with parameters
$\alpha$ = 1.51 $\pm$ 0.12, $\Phi^*$ = 0.006 $\pm$ 0.003,
and log M$^*$ = 9.7 $\pm$ 0.10.

\end{abstract}


\section{Introduction}

The obscuration due to dust and the high stellar density in our
Galaxy varies from place to place within the Milky Way.
Overall, it blocks our optical view of the extragalactic Universe
over $\sim$20\% of the sky, somewhat less in the infrared.
This ``Zone of Avoidance" (ZOA) was recognized even before the
nature of the spiral nebulae themselves was understood.
This sky coverage limitation
does not pose a problem for the study of galaxies themselves,
as there is no reason to believe that the population of obscured galaxies 
should differ from those in optically unobscured regions.
However, to understand the Local Group's motion
requires mapping the surrounding mass inhomogeneity, measured
in practice by galaxy over- and under-densities, across the entire sky.
In particular, the lack of a full census of nearby, hidden
galaxies is troublesome, since the local galaxies should play
a significant role in the Milky Way's motion with respect to the
microwave background (Kraan-Korteweg 1993). 

The ZOA has been successfully narrowed by deep searches in the
optical and infrared (see Kraan-Korteweg \& Woudt 1999 for a
comprehensive review of the various efforts).
However, optical searches fail where the extinction 
exceeds 4 - 5 magnitudes, within about $|b| \la 5^\circ$ of the
Galactic plane.
Near-infrared surveys, e.g. 2MASS (Skrutskie \etal 1997) and DENIS
(Epchtein 1997), will eventually produce catalogs of galaxies
closer to the plane than is possible with optical searches,
but they do not recover the most heavily obscured galaxies, or
galaxies of low surface brightness (Schr\"{o}der, Kraan-Korteweg,
\& Mamon 1999).
Far-infrared surveys become confusion limited by Galactic sources at low 
latitudes, and the remaining ZOA still covers $\sim10~\%$
of the sky.

Galaxies which contain \ion{H}{1} can be found in the regions of thickest
obscuration and IR confusion.
The technique was pioneered over a decade ago by Kerr \& Henning (1987)
who showed through a small, pilot survey that completely optically-hidden
galaxies could be readily uncovered through the detection of their
21-cm emission.
Since then, a spatially complete survey for spirals out to 4000 \kms has been
conducted over the northern ZOA ($30^\circ \le \ell \le 220^\circ$,
$|b|$~$\leq$~5$^\circ$; rms noise 40 mJy beam$^{-1}$)
with the 25~m Dwingeloo telescope (Henning \etal 1998, Rivers \etal 1999).
The survey uncovered no massive, ``Andromeda"-type galaxy in the ZOA, indeed the
nearest, previously unknown galaxy revealed by the survey was Dwingeloo
1, a likely member of the IC342/Maffei group (Kraan-Korteweg \etal 1994).
The census for nearby, \ion{H}{1}-bearing galaxies in the northern ZOA is
complete, at least for those galaxies whose redshifts or blueshifts
are sufficient to
separate their \ion{H}{1} signals from Galactic 21-cm emission, 
at $0$~$\pm$~$\sim$100 \kms.

We report here on a somewhat deeper survey (rms noise 15 mJy beam$^{-1}$)
for \HI galaxies in the southern
ZOA, conducted with the new multibeam receiver on the 64~m Parkes
radiotelescope
\footnote{The Parkes telescope is part of the Australia Telescope which 
is funded by the Commonwealth 
of Australia for operation as a National Facility
managed by CSIRO.} 
.
The angular coverage ($212^\circ \le \ell \le 36^\circ$,
$|b|$~$\leq$~5$^\circ$) completes the survey of the great circle of
the ZOA for relatively nearby, dynamically important HI galaxies.
The present survey discussed here (the ``shallow survey")
represents the first stage of an ongoing deeper search
of the area with the multibeam system.
The shallow survey is comprised of the first two scans of a planned
25 scans of the southern ZOA, estimated to be completed in mid-2000.
(This full sensitivity survey will be sensitive to spirals to
a redshift of $\sim$ 10,000 \kms).
In addition to the astronomical motivation outlined above, 
the shallow survey serves as a testbed of techniques
for the full sensitivity survey.

An intermediate-depth survey consisting of four scans of the region 
$308^\circ \le \ell \le 332^\circ$; $|b|$~$\leq$~5$^\circ$,
has been conducted (Juraszek \etal 2000).
This region is of particular interest as it contains the
predicted position of the core of the Great Attractor
(Kolatt, Dekel, \& Lahav 1995).

In \S~2, the observations and data reduction will be described.
The search method and galaxy \HI parametrization procedure
will be outlined in \S~3.
The resulting catalog is presented in \S~4.
Discussion of the galaxy distribution at low
Galactic latitudes, the \HI mass
function derived for the sample, and predictions for the full
sensitivity survey are contained in \S~5.

\section{Observations and Data Reduction}

The observations for the survey commenced on 1997 March 22 and were made by
scanning the Parkes multibeam receiver in strips of constant Galactic
latitude, each of length 8\arcdeg.  The Parkes receiver has 13
independent beams, each with two orthogonal linear polarizations. The receiver
rotation was fixed (with respect to the telescope which has an alt-az
mount) during each scan such that the rotation angle of the receiver,
relative to the scan direction, was 15\arcdeg~ at the mid-point of the
scan. This meant that the beams rotated on the sky during each
scan. Nevertheless, there was sufficient overlap to obtain complete
sampling of the entire southern Plane, with approximately uniform
sensitivity (see Staveley-Smith 1997).  The absence of continuous
receiver rotation and axial focussing helped ensure maximum baseline
stability. The telescope scan rate was 1 degree/min, so each
scan took 8 min to complete. The observations were done in parallel with the
\HI Parkes All Sky Survey (HIPASS) and a deeper ZOA survey (see
http://www.atnf.csiro.au/research/multibeam/ for current information regarding
these surveys).  Most observations for the shallow
survey were completed by 1997 September 4, although some regions were
re-scanned as recently as 1999 August 28, because of earlier pointing
problems in a few fields.

The footprint of the multibeam receiver on the sky is $\sim
1.7\arcdeg$, so that each scan maps out an $8\arcdeg \times \sim
1.7\arcdeg$ strip of longitude. To obtain full coverage of the
southern ZOA, 34 scans were made between Galactic latitudes
$-5\arcdeg$ and $+5\arcdeg$ at each of 23 separate central longitudes
from 216\arcdeg~ to 32\arcdeg. The observing parameters
and the area
mapped is summarized in Table 1.  In total 782 scans, or
$\sim104$ hrs of data, were obtained for the 1840 deg$^{2}$ comprising the
survey region. This corresponds to an integration time, when
integrated over all 13 beams, of 200 s beam$^{-1}$, where the beam
size, after gridding, has been taken to be 15\farcm5.
(The full-sensitivity southern ZOA survey will comprise 425 scans at each
longitude, or approximately 1300 hrs of integration time).

The average system temperature of the multibeam receiver during the
observations was $\sim25$ K. The average rms noise after Hanning
smoothing and away from strong continuum and line sources was 15 mJy
beam$^{-1}$. 
The velocity range
of the observations was $-1200$ to 12700 km s$^{-1}$.  The channel
spacing was 13.2 km~s$^{-1}$ and the resolution, after Hanning
smoothing, was 27.0 km ~s$^{-1}$.

The spectral data were bandpass-corrected, smoothed and Doppler-corrected
using {\sc aips++ LiveData} (Barnes 1998). 
For each beam and polarization, a 
template bandpass was calculated by taking the median of the surrounding data
with $\Delta \ell=\pm2\arcdeg$ and $\Delta t=\pm120$ s. The spectral data
were then gridded into 23 cubes, each of size 10\arcdeg~ in longitude by
8\arcdeg~ in latitude with a pixel size of 4\arcmin~.
Each output pixel consists of the median
of approximately 30 independent
(5 s) spectra which lie within a radius of 6\arcmin~ of the center of the pixel.
The use of the median statistic successfully removes the effect of outlyers,
so that
interference and bad data were rejected with good efficiency
without manual editing of all spectra (which number about $2\times
10^6$ for the shallow survey).
For more details, see Barnes \etal (2000).

\begin{deluxetable}{lc}
\scriptsize
\tablecaption{Shallow survey parameters}
\tablehead{
\colhead{Parameter} & \colhead{Value}  }
\startdata
& \\
Galactic longitude          & $212\arcdeg < \ell < 36\arcdeg$ \\
Galactic latitude           & $-5\arcdeg < b < 5\arcdeg$ \\
Velocity coverage           & $-1200 < cz < 12700$ km~s$^{-1}$ \\
Telescope FWHP resolution   & 14\farcm3 \\
FWHP resolution in cube        & 15\farcm5 \\
Integration time per beam   & 200 s \\
RMS noise\tablenotemark{a}  & 15 mJy beam$^{-1}$ \\
Velocity resolution\tablenotemark{a}~~(FWHP)  & 27.0 km~s$^{-1}$ \\
Approximate flux density limit & 75 mJy  beam$^{-1}$ \\
\enddata

\tablenotetext{a}{After Hanning smoothing.}
\end{deluxetable}

\section{Search Method and Profile Parametrization}

The cubes were Hanning smoothed, and each
of the 23 smoothed data cubes was searched by eye 
using the visualization tool {\sc karma kview} (Gooch 1995).
(Experiments with automatic galaxy detection algorithms have
failed to date in the complicated ZOA, mainly due
to baseline instabilities resulting from the presence of
strong Galactic continuum sources.
The eye/brain system is still far more effective for finding
the \HI signals).
First, Right Ascension -- velocity planes were examined, and
candidates noted.
The selection criterion was a minimum peak flux density of
$\sim$~75 mJy (5$\sigma$), with emission extended over two or more channels,
or a minimum flux integral of $\sim$~4 Jy \kms.
The candidates were then checked at the corresponding positions in
Right Ascension -- declination planes.
Candidates which were created by interference or increased noise
at the edges of fields were culled.
Finally, the one-dimensional profile, flux density versus velocity,
was created for each source, and its shape was checked for good 
sense, since \HI galaxies tend to have 2-horned, flat-topped, or
Gaussian profiles.

To measure central positions and flux integrals, a zeroth moment map was first
made by integrating the cubes in the full velocity range occupied by each 
galaxy. For most galaxies, which were unresolved, a Gaussian with the same 
FWHP diameter as the gridded telescope beam (15\farcm5) and a DC offset were 
simultaneously fitted to each zeroth moment map. Five galaxies were found to be 
significantly extended, and were fitted with Gaussians of arbitrary width.
The positional accuracy is 2\arcmin-3\arcmin, depending on the S/N ratio, and
the flux integral uncertainty is $\sim 20$\%.

Using the best-fit position, spectra were then extracted from the cubes
using the {\sc miriad}
task {\sc mbspect} (Sault, Teuben, \& Wright 1995). 
For the unresolved galaxies, 
this spectrum is a weighted average, with the weight dependent on the 
distance of each pixel from the fitted position. For resolved galaxies, 
the spectrum is spatially integrated across the galaxy. Velocity widths were
then measured at the 20\% and 50\% levels ($W_{20}$ and $W_{50}$), relative 
to the peak signal, using a width-maximizing algorithm (Lewis 1983). 
In order 
to correct for the relatively coarse velocity resolution (27 km s$^{-1}$) of 
the Hanning-smoothed spectra, downward corrections of 21 km s$^{-1}$ and
14 km s$^{-1}$ were applied to $W_{20}$ and $W_{50}$, respectively.
A central velocity was also measured from the 50\% values. All velocities 
quoted are in the optical ($cz$) convention.

\section{The Catalog}

The shallow survey detected 110 galaxies, whose \HI parameters are
listed in Table 2, and whose profiles are shown in Figure 1.
  \begin{figure}
  \begin{center}
  \psfig{file=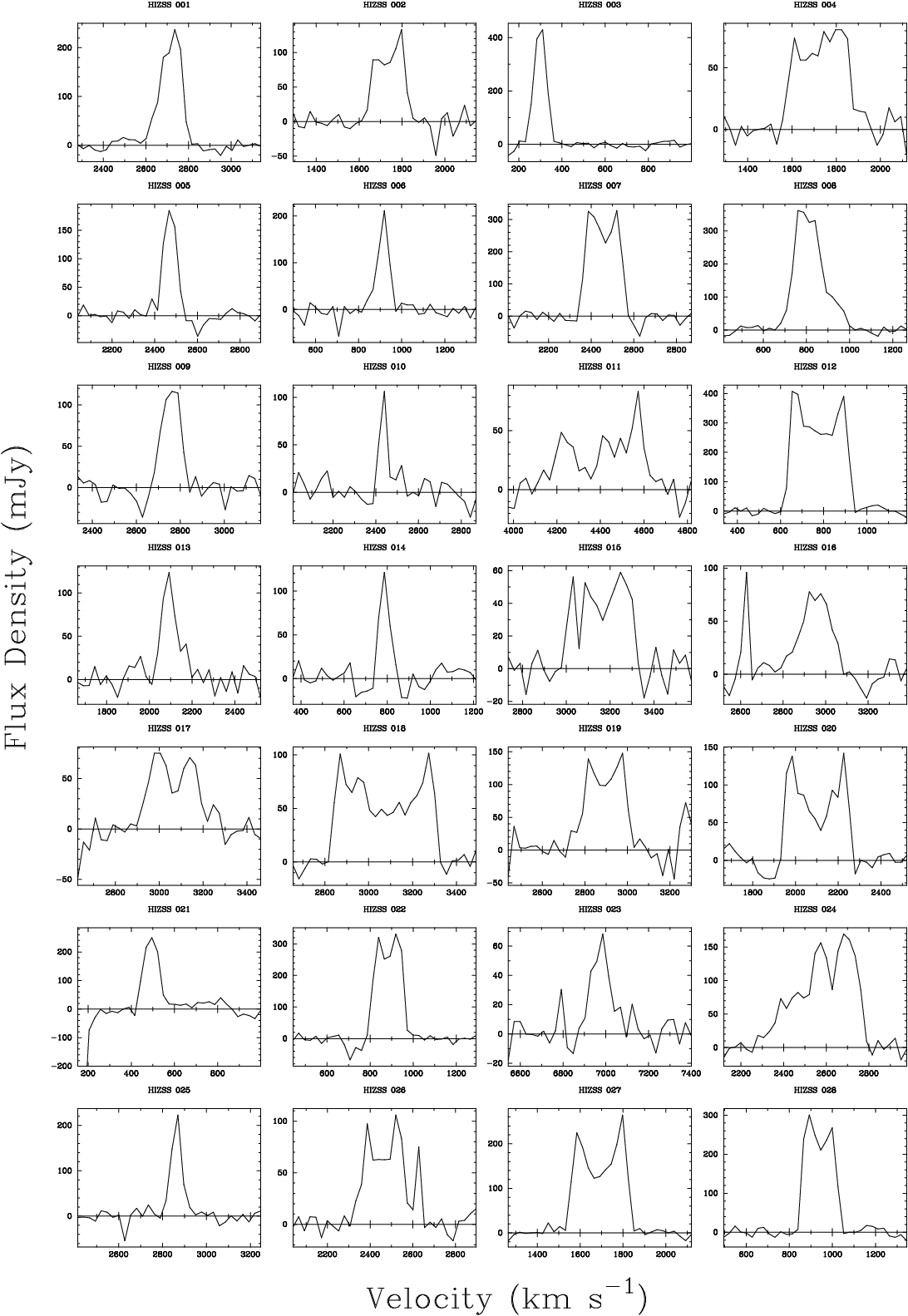,width=16cm,height=19cm}
  \end{center}
  \end{figure}
  \begin{figure}
  \begin{center}
  \psfig{file=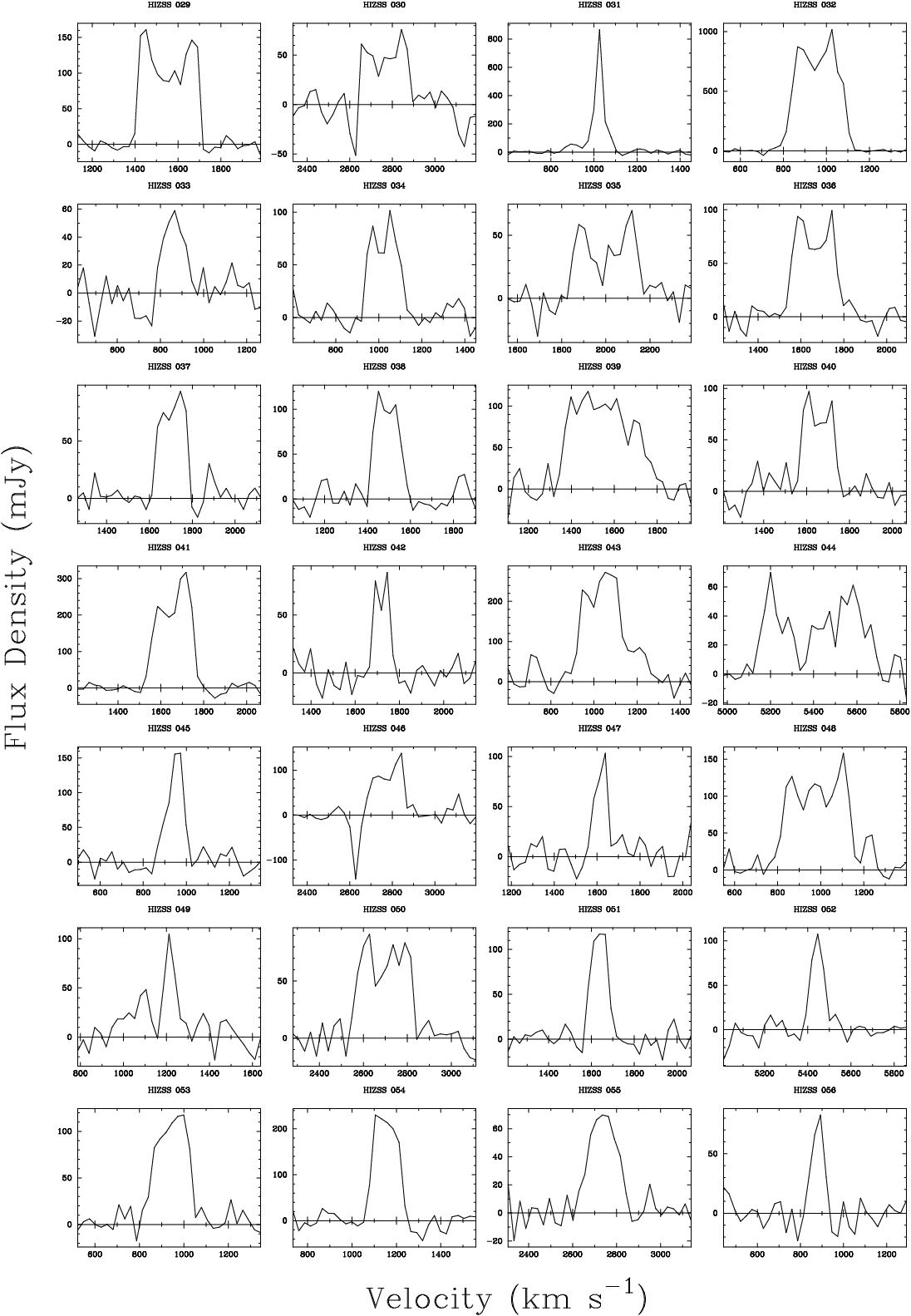,width=16cm,height=19cm}
  \end{center}
  \end{figure}
  \begin{figure}
  \begin{center}
  \psfig{file=henning.fig1_3.ps,width=16cm,height=19cm}
  \end{center}
  \end{figure}
  \begin{figure}
  \begin{center}
  \psfig{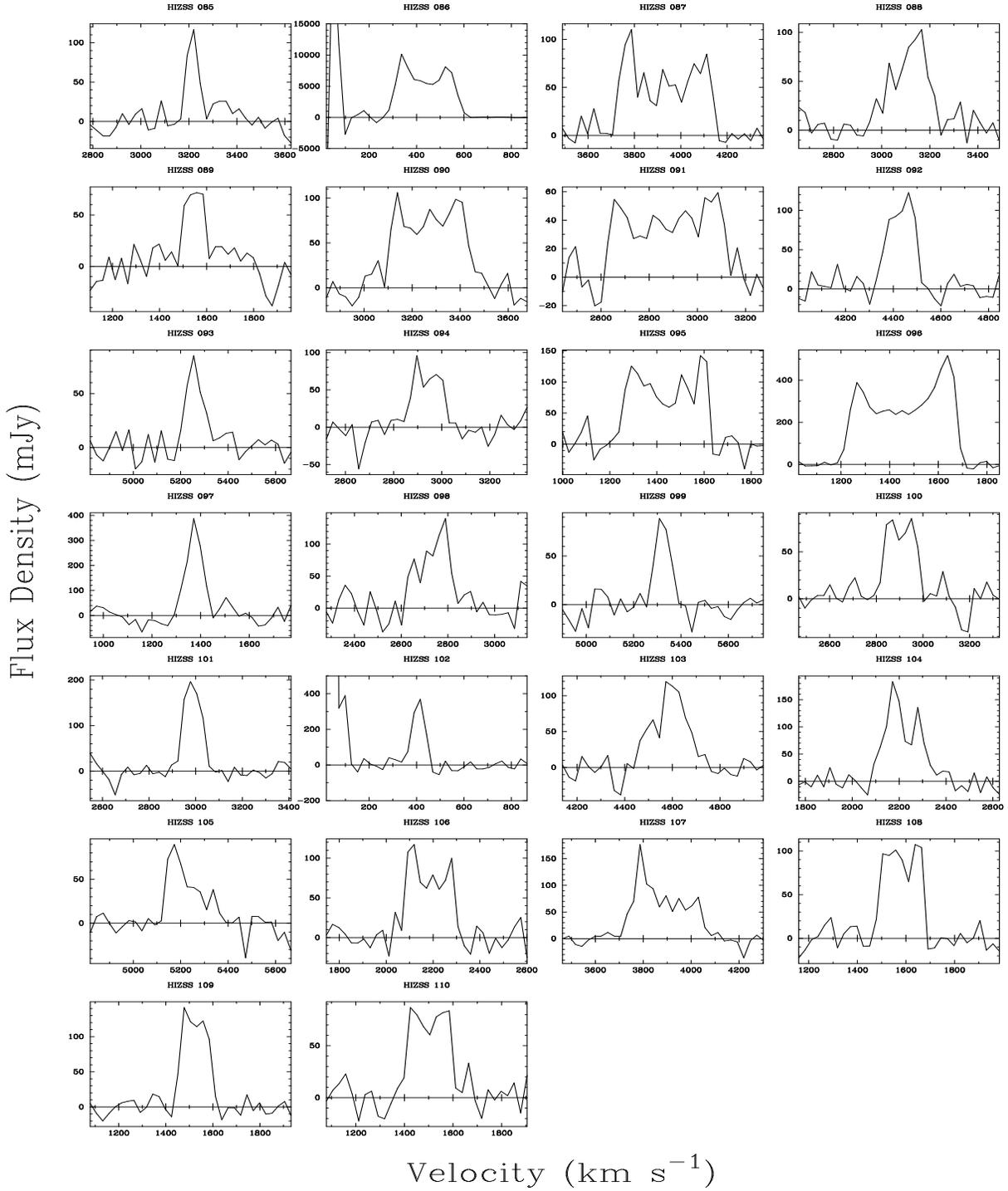}
  \caption{\HI profiles for the 110 shallow survey detections.
Baseline excursions near 0 \kms and 2650 \kms 
for HIZSS 016, 017, 021, 025, 026, 030, 046, 076,
077, 078, 086, 094, and 102 are due to
Galactic \HI and narrowband interference, respectively.}
  \label{fig-1}
  \end{center}
  \end{figure}
In Table 2, columns contain the following information:

Column 1a:  Source name;

Column 1b:  Indicates if the galaxy has an optical or IR counterpart
within 6 arcmin listed in
the NASA/IPAC Extragalactic Database (NED).
If the entry is followed by a colon, there is no redshift given in
NED for the object, or, for IRAS 13413-6525 (HIZSS 083)
and ESO 223-G012 (HIZSS 095), the HI redshift is just
outside of the uncertainty indicated in NED;

Columns 2a \& 2b:  Equatorial coordinates (J2000) of the fitted position;

Columns 3a \& 3b:  Galactic coordinates;

Column 4:  HI flux integral;

Column 5:  Heliocentric velocity ($cz$), the midpoint at 50\% of the profile's 
peak flux density;

Columns 6 \& 7:  Velocity width at 50\% of peak, and at 20\% of peak 
flux density;

Column 8:  Distance to the galaxy correcting the velocity to the Local
Group frame, and taking {\it H$_{0}$} to be 75 \kms~Mpc$^{-1}$;

Column 9: Logarithm of the \HI mass;

Column 10:  Foreground extinction A$_{\rm B}$ estimated from the IRAS/DIRBE
maps of Schlegel, Finkbeiner, \& Davis (1998).

\subsection{Completeness and Reliability}

Of these 110 galaxies, 29 have optical counterparts listed
in NED with matching redshifts.
Two more have counterparts in NED with redshifts just outside the
quoted accuracy.
A further 12 have a cataloged galaxy within 6 arcmin of the HI position, 
but with no redshift information.
The extinction at the positions of each galaxy was estimated from the
DIRBE/IRAS maps of Schlegel, Finkbeiner, \& Davis (1998).
Note that these maps are not calibrated this close to the Galactic
plane, so extinction measurements are somewhat uncertain.  
However, as expected,
the extinction is measured to be higher at the positions of the uncataloged
galaxies:  the median A$_{\rm B}$ for the cataloged galaxies is 2.9 mag, versus
5.7 at the positions of the uncataloged objects.

An intermediate-depth survey conducted with the multibeam system in a similar
way, but with four scans compared to the two discussed
here, of the region 
$308^\circ \le \ell \le 332^\circ$; $|b|$~$\leq$~5$^\circ$, detected
42 galaxies (Juraszek \etal 2000).
The intermediate-depth cubes and the shallow survey cubes
were searched independently.
All 24 of the shallow survey galaxies in this longitude range were recovered
by Juraszek \etal, as they should have been, owing to their $\sqrt{2}$
improved sensitivity.
Of the 18 other galaxies found by this deeper survey, 14 fell well below
the selection criteria of the shallow survey.
There were four, however, which Juraszek \etal determined to have
peak fluxes just at or above the shallow survey flux density cutoff
(J1414-62, J1417-55, J1526-51, and J1612-56; cf. Fig 2. in Juraszek \etal).
Two of these, J1414-62 and J1612-56, were noted as possible detections
in the course of examining the shallow survey cubes, but it was felt they
were not quite secure enough for inclusion in the present catalog,
the philosophy being that false detections are extremely undesirable.
The other two objects were missed.
One, J1417-55, while determined to be a broad-profiled galaxy
in the intermediate-depth survey, 
was not recognized as a galaxy by the present work, since only
one horn of the profile exceeds 75 mJy,
and so appears as a random spike, too narrow for inclusion in this
sample.
The other, J1526-51, peaks above 100 mJy in the intermediate-depth
survey, but has a linewidth of only W$_{50}$ = 21 \kms, and did not
satisfy the condition of appearing in at least two channels.
Thus the stated selection criteria of the shallow survey are consistent
with the results of the intermediate-depth survey in this region.

As a check on the accuracy of the measured parameters, the
21-cm properties of HIZSS 005 (IRAS 07112-0746), HIZSS 008 (IRAS 07232-2422),
HIZSS 018 (IRAS 07395-2224), HIZSS 024 (ESO 493-G016), HIZSS 027 (ESO 430-G001),
HIZSS 028 (ESO 561-G002), HIZSS 029 (ESO 494-G007), HIZSS 032 (ESO 494-G026),
HIZSS 034 (ESO 430-G020), HIZSS 036 (ESO 430-G026), HIZSS 037 (AM 0813-284),
HIZSS 038 (ESO 430-G030), HIZSS 039 (NGC 2559), HIZSS 040 (ESO 431-G001),
HIZSS 041 (UGCA 137), and HIZSS 044 (ESO 370-G015) were compared with
values found in the literature (Garcia \etal 1994,
Huchtmeier \& Richter 1986,
Kraan-Korteweg \& Huchtmeier 1992,
Staveley-Smith \& Davies 1987, 1988,
Takata \etal 1994, and Theureau \etal 1998).
There is excellent agreement amongst measurements of V$_{\rm hel}$,
W$_{20}$, and W$_{50}$, typically within a few \kms and only rarely 
differing by more than the velocity resolution.
There is more significant scatter amongst \HI flux integral measurements,
reflecting the measurement uncertainty of the shallow survey ($\sim20\%$)
and in literature values.
The HIZSS flux integrals are neither systematically higher nor lower than
previously published values.

\begin{deluxetable}{lrrrrrrrrrlcc}
\tablecolumns{10}
  \scriptsize
\tablecaption{Galaxy Parameters from HI Parkes Multibeam Observations}
  \tablehead{
\colhead{Name} & \colhead{R.A.} & \colhead{l} &
\colhead{Flux Int} & \colhead{V$_{\rm hel}$} & \colhead{W$_{50}$} & \colhead{W$_{20}$} &
\colhead{Dist} & \colhead{log M$_{\rm HI}$} & \colhead{A$_{\rm B}$} \\
\colhead{Opt/IR} & \colhead{Dec} & \colhead{b} & 
\colhead{} & \colhead{} & \colhead{} & \colhead{} & \colhead{} &
\colhead{} & \colhead{} \\
\colhead{} & \colhead{(J2000)} & \colhead{\arcdeg} & \colhead{Jy \kms} &
\colhead{} & \colhead{\kms} & \colhead{} & \colhead{Mpc} & \colhead{M$_\odot$} &
\colhead{mag}\\
}
\startdata
HIZSS 001 &06 57 57&218.42&  19.2&2720& 98&145&33.8& 9.7& 5.5\nl
IRAS 06555-0516&$-$05 20 04& $-$0.99\nl
\nl
HIZSS 002 &06 59 24&215.18&  16.7&1736&147&173&20.8& 9.2& 3.3\nl
CGMW 1-0476:&$-$01 30 15& 1.08\nl
\nl
HIZSS 003 &07 00 26&217.70&  32.1& 299& 55& 85& 1.5& 7.3& 4.7\nl
--&$-$04 12 17& 0.08\nl
\nl
HIZSS 004 &07 09 18&219.80&  20.4&1725&267&297&20.4& 9.3& 2.0\nl
IRAS 07071-0520&$-$05 25 54& 1.48\nl
\nl
HIZSS 005 &07 13 33&222.45&  12.1&2470& 62& 83&30.2& 9.4& 2.7\nl
IRAS 07112-0746&$-$07 51 51& 1.29\nl
\nl
HIZSS 006 &07 18 16&224.07&  12.0& 915& 41& 72& 9.4& 8.4& 1.9\nl
--&$-$09 04 43& 1.76\nl
\nl
HIZSS 007 &07 24 54&225.35&  45.7&2457&168&193&29.9&10.0& 1.5\nl
NGC 2377&$-$09 39 51& 2.93\nl
\nl
HIZSS 008 &07 25 14&238.45&  56.3& 805&126&222& 7.3& 8.9& 6.1\nl
IRAS 07232-2422&$-$24 28 25&$-$4.01\nl
\nl
HIZSS 009 &07 25 47&232.70&  11.3&2756& 92&121&33.6& 9.5& 6.1\nl
IRAS 07235-1747:&$-$17 53 14&$-$0.78\nl
CGMW 1-0879:&\nl 
\nl
HIZSS 010 &07 26 38&225.17&   5.1&2435& 27& 43&29.6& 9.0& 1.4\nl
ZOAG G225.17+03.49:&$-$09 13 24& 3.51\nl
\nl
HIZSS 011 &07 27 39&238.25&  14.3&4404&371&423&55.3&10.0& 5.3\nl
CGMW 2-0889:&$-$23 57 17&$-$3.28\nl
\nl
HIZSS 012 &07 30 08&236.80&  83.7& 776&264&284& 7.0& 9.0& 7.9\nl
--&$-$22 00 11&$-$1.84\nl
\nl
HIZSS 013 &07 33 10&242.99&   8.6&2089& 60&127&24.3& 9.1& 2.2\nl
CGMW 2-0978:&$-$28 39 41&$-$4.43\nl
\nl
HIZSS 014 &07 36 09&235.25&   3.4& 785& 41& 68& 7.2& 7.6& 5.7\nl
--&$-$19 27 23& 0.62\nl
\nl
HIZSS 015 &07 38 02&233.29&  14.5&3157&291&313&38.9& 9.7& 2.8\nl
IRAS 07357-1651&$-$16 57 04& 2.23\nl
\nl
HIZSS 016 &07 38 55&241.47&  11.5&2957&146&237&35.9& 9.5& 3.7\nl
CGMW 2-1059:&$-$26 13 12&$-$2.13\nl
CGMW 2-1056:&\nl
\nl
HIZSS 017 &07 40 05&240.27&  12.3&3062&229&274&37.4& 9.6& 3.4\nl
IRAS 07377-2435:&$-$24 41 40&$-$1.15\nl
\nl
HIZSS 018 &07 41 47&238.56&  29.7&3073&451&472&37.6&10.0& 2.9\nl
IRAS 07395-2224&$-$22 30 47& 0.26&\nl
\nl
HIZSS 019 &07 42 34&249.19&  24.9&2897&191&233&34.9& 9.9& 6.1\nl
--&$-$34 37 11&$-$5.57\nl
\nl
HIZSS 020 &07 42 48&246.89&  26.0&2099&289&305&24.3& 9.6& 3.5\nl
CGMW 2-1147:&$-$31 57 39&$-$4.22\nl
\nl
HIZSS 021 &07 46 06&244.17&  25.4& 491& 72& 95& 2.9& 7.7& 4.3\nl
--&$-$28 25 35&$-$1.84\nl
\nl
HIZSS 022 &07 46 57&242.46&  37.5& 884&133&152& 8.2& 8.8& 2.8\nl
--&$-$26 20 01&$-$0.63\nl
\nl
HIZSS 023 &07 47 21&238.44&   7.3&6971& 82&148&89.5&10.1& 2.6\nl
CGMW 2-1245&$-$21 37 34& 1.82\nl
\nl
HIZSS 024 &07 48 39&242.56&  36.2&2642&225&405&31.7& 9.9& 2.9\nl
ESO 493-G016&$-$26 13 41&$-$0.25\nl
\nl
HIZSS 025 &07 49 38&250.84&  11.9&2861& 41& 70&34.4& 9.5& 6.3\nl
--&$-$35 41 28&$-$4.84\nl
\nl
HIZSS 026 &07 53 45&245.66&  17.8&2463&180&221&29.2& 9.6& 3.9\nl
IRAS 07517-2903:&$-$29 10 03&$-$0.78\nl
\nl
HIZSS 027 &07 55 10&244.96&  48.5&1692&246&276&18.9& 9.6& 3.7\nl
ESO 430-G001&$-$28 09 17& 0.01\nl
\nl
HIZSS 028 &07 55 24&239.15&  36.6& 938&149&171& 9.1& 8.9& 1.5\nl
ESO 561-G002&$-$21 20 25& 3.58\nl
\nl
HIZSS 029 &07 56 51&242.36&  28.9&1556&276&288&17.2& 9.3& 1.4\nl
ESO 494-G007&$-$24 53 15& 2.03\nl
\nl
HIZSS 030 &08 01 44&246.50&   9.3&2764&215&225&33.2& 9.4& 4.1\nl
--&$-$29 05 05& 0.76\nl
\nl
HIZSS 031 &08 05 22&245.46&  30.4&1024& 24& 57&10.0& 8.9& 1.6\nl
--&$-$27 21 13& 2.35\nl
\nl
HIZSS 032 &08 06 12&245.70& 197.4& 961&228&264& 9.2& 9.6& 1.8\nl
ESO 494-G026&$-$27 31 16& 2.42\nl
\nl
HIZSS 033 &08 07 02&254.42&   4.6& 862&108&138& 7.6& 7.8& 5.7\nl
--&$-$37 44 47&$-$2.93\nl
\nl
HIZSS 034 &08 07 09&246.24&  12.8&1022&147&173&10.0& 8.5& 2.0\nl
ESO 430-G020&$-$28 01 19& 2.33\nl
\nl
HIZSS 035 &08 09 58&258.05&  11.3&1998&281&307&22.7& 9.1& 5.2\nl
--&$-$41 41 30&$-$4.59\nl
\nl
HIZSS 036 &08 14 45&249.93&  13.4&1660&198&230&18.4& 9.0& 2.4\nl
ESO 430-G026&$-$31 20 54& 1.89\nl
\nl
HIZSS 037 &08 15 55&248.00&   9.3&1705&135&150&19.0& 8.9& 1.7\nl
AM 0813-284&$-$28 51 25& 3.49\nl
\nl
HIZSS 038 &08 16 31&247.43&  14.7&1488&123&149&16.2& 9.0& 1.5\nl
ESO 430-G030&$-$28 05 44& 4.02\nl
\nl
HIZSS 039 &08 17 12&246.97&  28.6&1548&353&414&17.0& 9.3& 0.9\nl
NGC 2559&$-$27 26 06& 4.51\nl
\nl
HIZSS 040 &08 17 41&248.95&  11.2&1653&147&165&18.3& 8.9& 1.4\nl
ESO 431-G001&$-$29 44 34& 3.31\nl
\nl
HIZSS 041 &08 17 43&249.26&  47.2&1659&174&206&18.4& 9.6& 1.7\nl
UGCA 137&$-$30 06 54& 3.11\nl
\nl
HIZSS 042 &08 20 57&251.92&   4.8&1719& 68& 80&19.1& 8.6& 1.7\nl
--&$-$32 51 49& 2.13\nl
\nl
HIZSS 043 &08 26 28&261.93&  41.6&1029&183&314& 9.8& 9.0& 6.5\nl
--&$-$44 19 23&$-$3.56\nl
\nl
HIZSS 044 &08 28 38&256.35&  18.6&5393&448&533&68.0&10.3& 3.7\nl
ESO 370-G015&$-$37 10 30& 0.93\nl
\nl
HIZSS 045 &08 34 12&257.32&  12.7& 952& 65&114& 8.8& 8.4& 3.8\nl
--&$-$37 33 48& 1.60\nl
\nl
HIZSS 046 &08 34 41&259.45&   8.4&2779&143&168&33.1& 9.3& 9.2\nl
--&$-$40 08 38& 0.13\nl
\nl
HIZSS 047 &08 55 32&269.47&   7.3&1616& 57& 76&17.6& 8.7& 7.2\nl
--&$-$50 01 38&$-$3.14\nl
\nl
HIZSS 048 &08 57 25&261.50&  31.8& 982&297&332& 9.1& 8.8& 3.2\nl
ESO 314-G002:&$-$39 16 24& 4.09\nl
\nl
HIZSS 049 &08 58 04&261.46&   6.3&1216& 44& 72&12.3& 8.3& 2.7\nl
--&$-$39 07 29& 4.28\nl
\nl
HIZSS 050 &08 58 32&266.18&  17.3&2696&244&266&32.0& 9.6&16.1\nl
--&$-$45 15 39& 0.34\nl
\nl
HIZSS 051 &09 01 53&262.72&  12.2&1634& 83&107&17.8& 9.0& 2.4\nl
--&$-$40 08 28& 4.17\nl
\nl
HIZSS 052 &09 03 30&263.78&   7.1&5444& 58& 80&68.6& 9.9& 4.5\nl
--&$-$41 17 13& 3.64\nl
\nl
HIZSS 053 &09 17 42&274.26&  20.6& 944&165&196& 8.6& 8.6& 3.6\nl
--&$-$53 22 30&$-$2.85\nl
\nl
HIZSS 054 &09 27 58&277.16&  27.9&1153&122&146&11.4& 8.9& 3.9\nl
--&$-$55 59 26&$-$3.66\nl
\nl
HIZSS 055 &09 37 06&277.19&   8.6&2741&145&195&32.6& 9.3& 8.4\nl
--&$-$54 37 22&$-$1.77\nl
\nl
HIZSS 056 &09 45 38&273.92&   3.1& 880& 52& 78& 7.8& 7.6& 2.0\nl
--&$-$48 07 27& 4.00\nl
\nl
HIZSS 057 &09 48 01&278.40&   8.1&1734&100&114&19.2& 8.8&10.6\nl
--&$-$54 39 25&$-$0.77\nl
\nl
HIZSS 058 &09 49 12&274.31&  20.5&1934&206&248&21.8& 9.4& 1.5\nl
ESO 213-G002&$-$48 01 00& 4.47\nl
\nl
HIZSS 059 &09 49 43&279.79&  30.7&1759&203&236&19.5& 9.4& 9.2\nl
--&$-$56 32 08&$-$2.06\nl
\nl
HIZSS 060 &09 57 14&275.89&  11.7&3725& 34& 57&45.7& 9.8& 2.5\nl
--&$-$48 52 26& 4.63\nl
\nl
HIZSS 061 &10 04 16&282.57&  15.6&3700&225&302&45.4& 9.9&15.2\nl
--&$-$58 33 35&$-$2.46\nl
\nl
HIZSS 062 &10 13 48&280.05&  18.8&2707&221&258&32.2& 9.7& 2.1\nl
ESO 213-G009&$-$52 18 14& 3.42\nl
\nl
HIZSS 063 &10 24 59&282.80&  10.1&1083&115&175&10.5& 8.4& 2.7\nl
ESO 168-G002&$-$54 47 13& 2.26\nl
\nl
HIZSS 064 &10 37 24&284.40&  15.8&2670&310&350&31.7& 9.6& 3.1\nl
ESO 168-G009:&$-$54 54 39& 3.06\nl
\nl
HIZSS 065 &10 39 51&284.52&  12.8&2757&105&127&32.9& 9.5& 2.8\nl
--&$-$54 31 32& 3.57\nl
\nl
HIZSS 066 &10 53 42&289.95&  32.3&1835&249&263&20.7& 9.5& 3.6\nl
--&$-$62 50 30&$-$2.98\nl
\nl
HIZSS 067 &11 30 39&292.61&  13.1&1841&131&152&20.9& 9.1& 3.8\nl
--&$-$58 46 04& 2.48\nl
\nl
HIZSS 068 &11 41 14&295.46&  41.4&2025&171&199&23.4& 9.7& 7.9\nl
--&$-$64 29 02&$-$2.64\nl
\nl
HIZSS 069 &11 49 44&296.23&  62.0&2114&157&312&24.6& 9.9&10.5\nl
--&$-$64 00 34&$-$1.94\nl
\nl
HIZSS 070 &12 02 45&297.19&  96.9&1540&202&224&17.0& 9.8&11.2\nl
--&$-$61 40 29& 0.65\nl
\nl
HIZSS 071 &12 04 16&297.65&  32.6&2042&167&211&23.7& 9.6&24.8\nl
--&$-$63 13 10&$-$0.83\nl
\nl
HIZSS 072 &12 13 37&299.03&  16.5&2144&223&262&25.1& 9.4& 3.7\nl
--&$-$65 33 38&$-$2.98\nl
\nl
HIZSS 073 &12 21 31&299.16&  44.4&1476&176&190&16.2& 9.4& 5.8\nl
--&$-$59 42 40& 2.94\nl
\nl
HIZSS 074 &12 45 46&302.29&  27.6&3917&449&473&48.9&10.2&24.6\nl
--&$-$63 05 27&$-$0.23\nl
\nl
HIZSS 075 &13 02 26&304.13&  13.7&3941& 68&214&49.2& 9.9&12.7\nl
--&$-$64 08 15&$-$1.29\nl
\nl
HIZSS 076 &13 12 49&305.53&  16.8&2317&214&238&27.6& 9.5& 9.8\nl
--&$-$60 53 42& 1.87\nl
\nl
HIZSS 077 &13 14 50&305.95&  29.1&2337&131&193&27.9& 9.7& 4.0\nl
WKK 2029&$-$58 55 47& 3.80\nl
\nl
HIZSS 078 &13 27 32&307.78&  24.4&2915&267&326&35.7& 9.9& 3.5\nl
ESO 173-G015&$-$57 31 04& 5.01\nl
\nl
HIZSS 079 &13 29 56&307.46&  18.7&3694&200&235&46.1&10.0&12.5\nl
--&$-$61 50 46& 0.69\nl
\nl
HIZSS 080 &13 33 07&308.43&  14.3&1482&107&132&16.6& 9.0& 3.5\nl
--&$-$58 06 09& 4.32\nl
\nl
HIZSS 081 &13 37 06&308.88&  12.6&4124&155&220&51.9& 9.9& 3.8\nl
--&$-$58 31 04& 3.83\nl
\nl
HIZSS 082 &13 42 14&309.04&  14.3&3942&284&355&49.5& 9.9&12.0\nl
--&$-$61 04 58& 1.19\nl
\nl
HIZSS 083 &13 45 15&308.44&  30.2&2789&292&359&34.1& 9.9& 3.9\nl
IRAS 13413-6525:&$-$65 39 25&$-$3.36\nl
\nl
HIZSS 084 &13 51 44&310.72&  23.8&3869&621&698&48.6&10.1& 4.5\nl
--&$-$58 39 44& 3.30\nl
\nl
HIZSS 085 &14 07 38&312.70&   4.7&3213& 44& 68&39.9& 9.2& 8.7\nl
--&$-$58 43 59& 2.69\nl
\nl
HIZSS 086 &14 13 27&311.36&1866.6& 436&240&278& 2.8& 9.5& 6.1\nl
Circinus&$-$65 19 09&$-$3.80\nl
\nl
HIZSS 087 &14 19 42&314.37&  16.2&3931&385&413&49.6&10.0& 6.6\nl
--&$-$58 09 58& 2.73\nl
\nl
HIZSS 088 &14 32 07&316.92&  11.4&3109&160&244&38.7& 9.6& 3.7\nl
ESO 175-G009&$-$55 29 03& 4.63\nl
\nl
HIZSS 089 &14 36 12&314.48&   7.6&1546& 91&103&17.8& 8.8& 4.6\nl
--&$-$63 03 58&$-$2.57\nl
\nl
HIZSS 090 &14 52 57&319.03&  19.9&3269&311&344&41.0& 9.9& 6.9\nl
--&$-$56 46 07& 2.26\nl
\nl
HIZSS 091 &14 57 07&320.64&  23.9&2875&470&489&35.8& 9.9& 3.7\nl
ESO 176-G006:&$-$54 24 04& 4.09\nl
\nl
HIZSS 092 &15 01 27&318.17&  14.6&4433&121&153&56.4&10.0& 8.4\nl
--&$-$60 43 22&$-$1.76\nl
\nl
HIZSS 093 &15 04 15&321.04&   6.2&5256& 62&103&67.6& 9.8& 6.1\nl
--&$-$55 27 49&2.67\nl
\nl
HIZSS 094 &15 05 54&321.77&   8.7&2942&123&151&36.8& 9.4& 4.2\nl
--&$-$54 23 34& 3.49\nl
\nl
HIZSS 095 &15 09 28&323.15&  32.5&1440&349&367&16.8& 9.3& 3.3\nl
ESO 223-G012:&$-$52 33 27& 4.81\nl
\nl
HIZSS 096 &15 14 36&323.58& 155.4&1457&425&451&17.1&10.0& 4.3\nl
IRAS 15109-5248&$-$53 00 55& 4.02\nl
\nl
HIZSS 097 &15 32 27&324.10&  58.5&1375& 58&100&16.0& 9.5&78.0\nl
--&$-$56 01 56& 0.08\nl
\nl
HIZSS 098 &15 36 51&324.36&  14.6&2729&148&189&34.1& 9.6&31.2\nl
--&$-$56 26 19&$-$0.61\nl
\nl
HIZSS 099 &15 40 17&328.08&   6.0&5322& 62& 91&68.9& 9.8& 3.7\nl
--&$-$50 51 35& 3.58\nl
\nl
HIZSS 100 &15 43 10&323.66&   9.6&2905&142&158&36.4& 9.5& 5.0\nl
--&$-$58 44 24&$-$2.96\nl
\nl
HIZSS 101 &16 05 22&326.47&  18.5&2987& 83&102&37.6& 9.8& 2.1\nl
--&$-$57 51 43&$-$4.15\nl
\nl
HIZSS 102 &16 18 34&329.31&  30.3& 409& 50& 71& 3.4& 7.9& 2.7\nl
--&$-$55 38 44&$-$3.77\nl
\nl
HIZSS 103 &16 20 09&336.01&  16.7&4586&147&225&59.5&10.1& 9.4\nl
--&$-$46 21 41& 2.67\nl
\nl
HIZSS 104 &16 24 33&339.32&  21.1&2218&143&210&28.2& 9.6& 4.8\nl
--&$-$42 29 05& 4.85\nl
\nl
HIZSS 105 &16 53 05&346.34&  15.1&5179& 75&212&68.1&10.2& 6.0\nl
--&$-$37 57 23& 3.79\nl
\nl
HIZSS 106 &17 11 46&340.80&  16.1&2185&199&213&27.8& 9.5& 2.9\nl
--&$-$47 35 24&$-$4.81\nl
\nl
HIZSS 107 &17 19 42&346.77&  25.6&3804& 66&304&49.8&10.2&11.4\nl
--&$-$41 17 18&$-$2.31\nl
\nl
HIZSS 108 &18 55 55& 30.57&  13.7&1582&172&184&23.1& 9.2& 6.0\nl
--&$-$03 13 27&$-$2.47\nl
\nl
HIZSS 109 &19 01 48& 30.10&  13.7&1525&120&142&22.3& 9.2& 3.5\nl
--&$-$04 29 51&$-$4.35\nl
\nl
HIZSS 110 &19 10 22& 35.55&  14.0&1503&177&193&22.4& 9.2& 2.9\nl
--&$+$00 30 32&$-$3.98\nl
\enddata
\end{deluxetable}
\section{Discussion}

\subsection{Nearby, \HI Bright Galaxies in the Southern ZOA}

The shallow survey's rms noise of 15 mJy is equivalent to a
5$\sigma$ HI mass detection limit of $4 \times 10^6$ d$^2\!\!_{\rm Mpc}$
M$_{\odot}$ (for a galaxy with the typical linewidth of
200 \kms).
Thus, the sensitivity to normal spirals falls rapidly beyond about
40 Mpc, and the survey is not well suited to discuss large-scale structure
behind the southern Milky Way beyond a few tens of Mpc.
The full survey will be sensitive to spirals to a much larger
redshift, $\sim$10,000 \kms, and will be able to address issues
such as the Great Attractor, predicted to lie at (l, b, v) $\sim$
(320$^\circ$, 0$^\circ$, 4500 \kms) (Kolatt, Dekel, \& Lahav 1995),
and other features which may lie hidden (see Kraan-Korteweg \& Woudt 1999
for a review of these structures, and Kraan-Korteweg \& Juraszek 2000
for preliminary analysis of multibeam 
full-sensitivity survey detections in the Great
Attractor region).
However, the shallow survey does now complete the census of nearby,
HI bright galaxies which lie at redshifts away from Galactic HI.
Sixteen of the 110 galaxies lie at redshifts less than 1000~\kms
and are, therefore, fairly nearby.
However, all of the shallow survey galaxies have HI flux integrals
an order of magnitude or more below that of the Circinus galaxy, 
a nearby, massive, low-Galactic latitude galaxy,
which is cataloged here as HIZSS 086.
One can estimate the dynamical importance of each hidden
galaxy without knowledge of its inclination or
precise distance by realizing that if
the galaxies in the catalog have similar
M$_{\rm HI}$ / M$_{\rm total}$,
the gravitational force on the Milky Way due to one of these galaxies
is proportional to its HI flux integral.
The closest rival in the catalog
to Circinus is ESO 494-G026 (Lauberts 1982), listed here
as HIZSS 032.
It has only about 10\% of Circinus's HI flux.
(Dynamical analysis of this object to estimate its total mass will
be conducted using HI synthesis data obtained with the VLA).
All of the newly discovered galaxies have even lower HI flux integrals. 
No single, previously unknown, dynamically important HI galaxy was found.
Five of the 110 galaxies are extended objects at the multibeam
resolution, and of these, three
were previously cataloged:  Circinus, ESO 494-G026, and
IRAS 15109-5248.
21-cm follow-up synthesis mapping has been done for IRAS 15109-5248 and
the two uncataloged, extended objects, HIZSS 097 and HIZSS 102 by
Staveley-Smith \etal (1998).
Rotation curve analysis indicates the IRAS galaxy is a massive disk system,
estimated M$_{\rm tot}$ = 6~$\times$ 10$^{11}$ M$_\odot$.
The other two are moderate to low mass systems, which under the
higher resolution scrutiny of the synthesis observations (synthesized
beam FWHM $\sim$ 2.2 - 4.5 arcmin) were seen to break up into apparently
interacting systems of low \HI column density.
\HI synthesis observations of all of the shallow survey detections
using the Australia Telescope
Compact Array and the NRAO Very Large Array have been completed or are
planned.
Identifications on the currently available strips of the southern
sky NIR survey DENIS (DEep NIR southern sky Survey; Epchtein 1997)
show that many of the shallow survey sources are visible on
the NIR images (Schr\"{o}der, Kraan-Korteweg, \& Mamon 1999;
Schr\"{o}der \etal in prep.).

\subsection{Large-Scale Structure in the Shallow Survey}

Figure 2 shows the distribution on the sky of the shallow survey detections,
and galaxies from the literature within a redshift of 4000 \kms.
  \begin{figure}
  \psfig{file=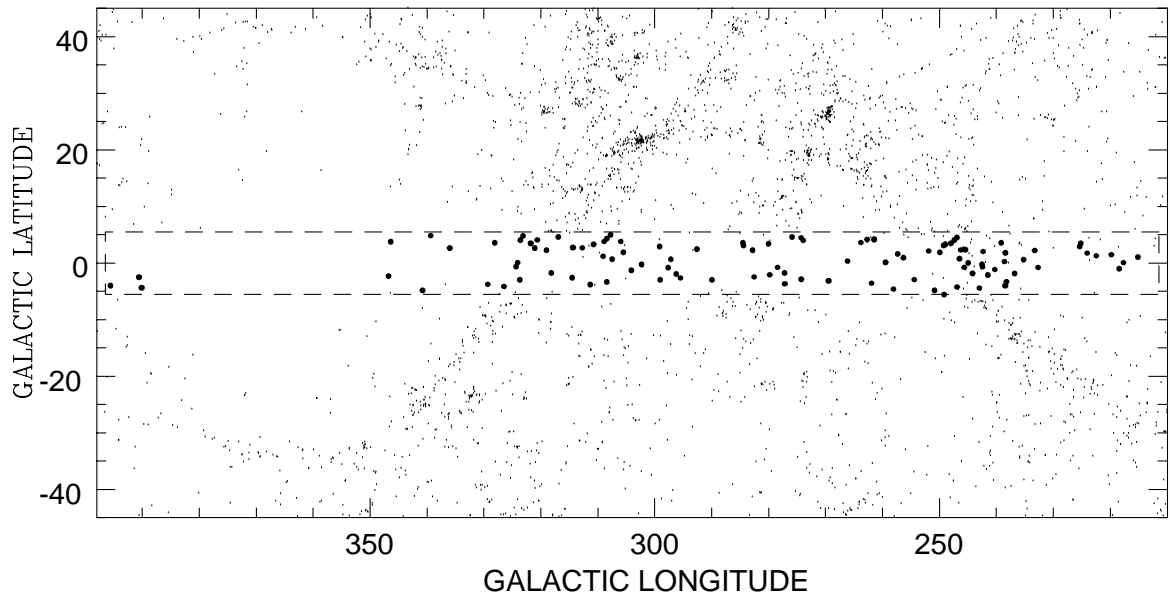,width=17cm}
  \caption{Distribution on the sky of cataloged galaxies from the LEDA
database within 4000 \kms
(small dots) and shallow survey detections (larger dots).
The shallow survey search area is delineated by the dashed rectangle.}
  \label{fig-2}
  \end{figure}
The shallow survey now fills in the southern ZOA within about 40 Mpc
for the first time, with minimal dependence of detection-rate on 
Galactic latitude (Fig. 3).
  \begin{figure}
  \psfig{file=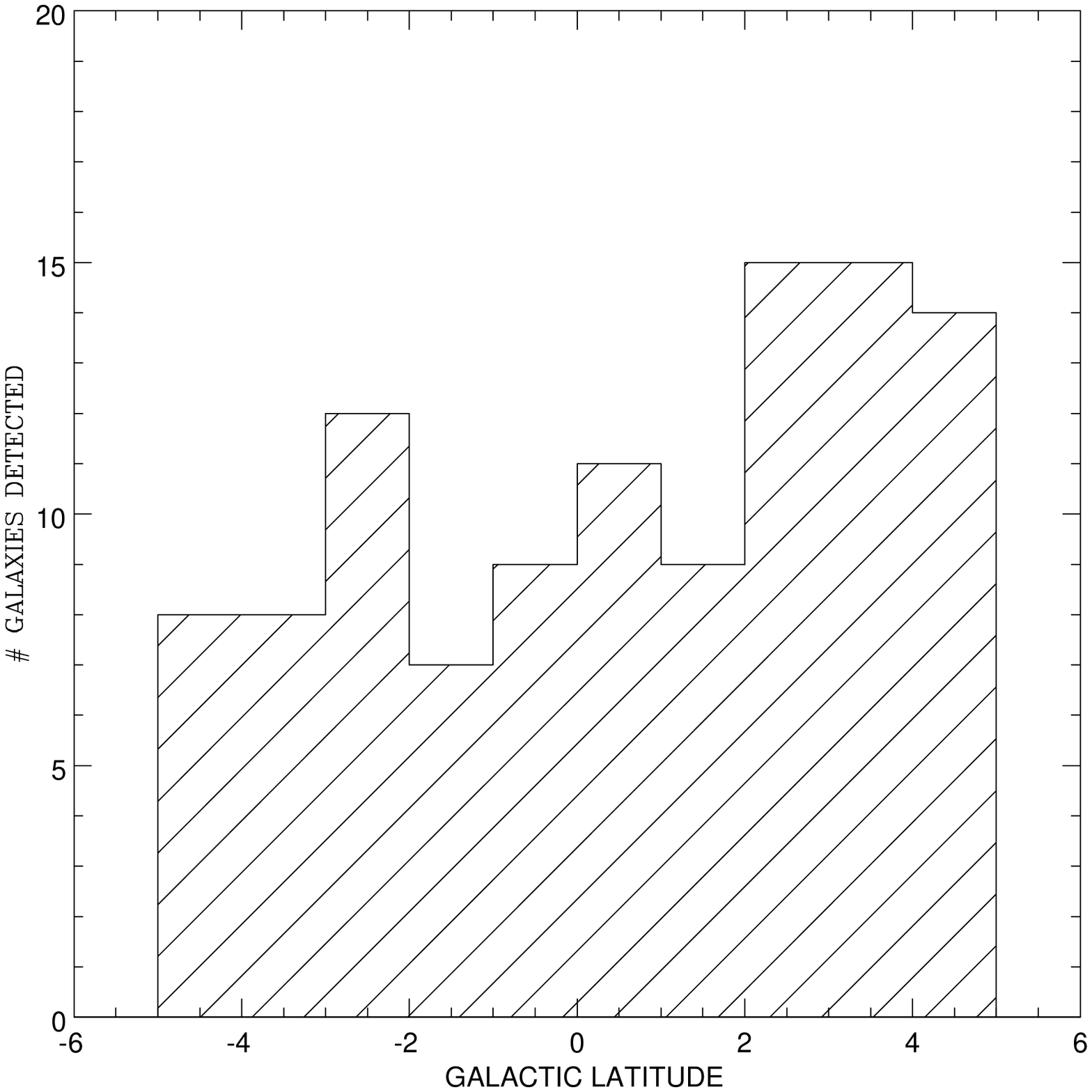,width=17cm}
  \caption{Number of shallow survey detections as a function of Galactic
latitude.}
  \label{fig-3}
  \end{figure}
The Local Void is clearly evident in the distribution of both the 
optically-cataloged objects, and the shallow survey detections. 
It is also apparent in Figure 4, the longitude-velocity distribution
of the HI detected objects.
  \begin{figure}
  \psfig{file=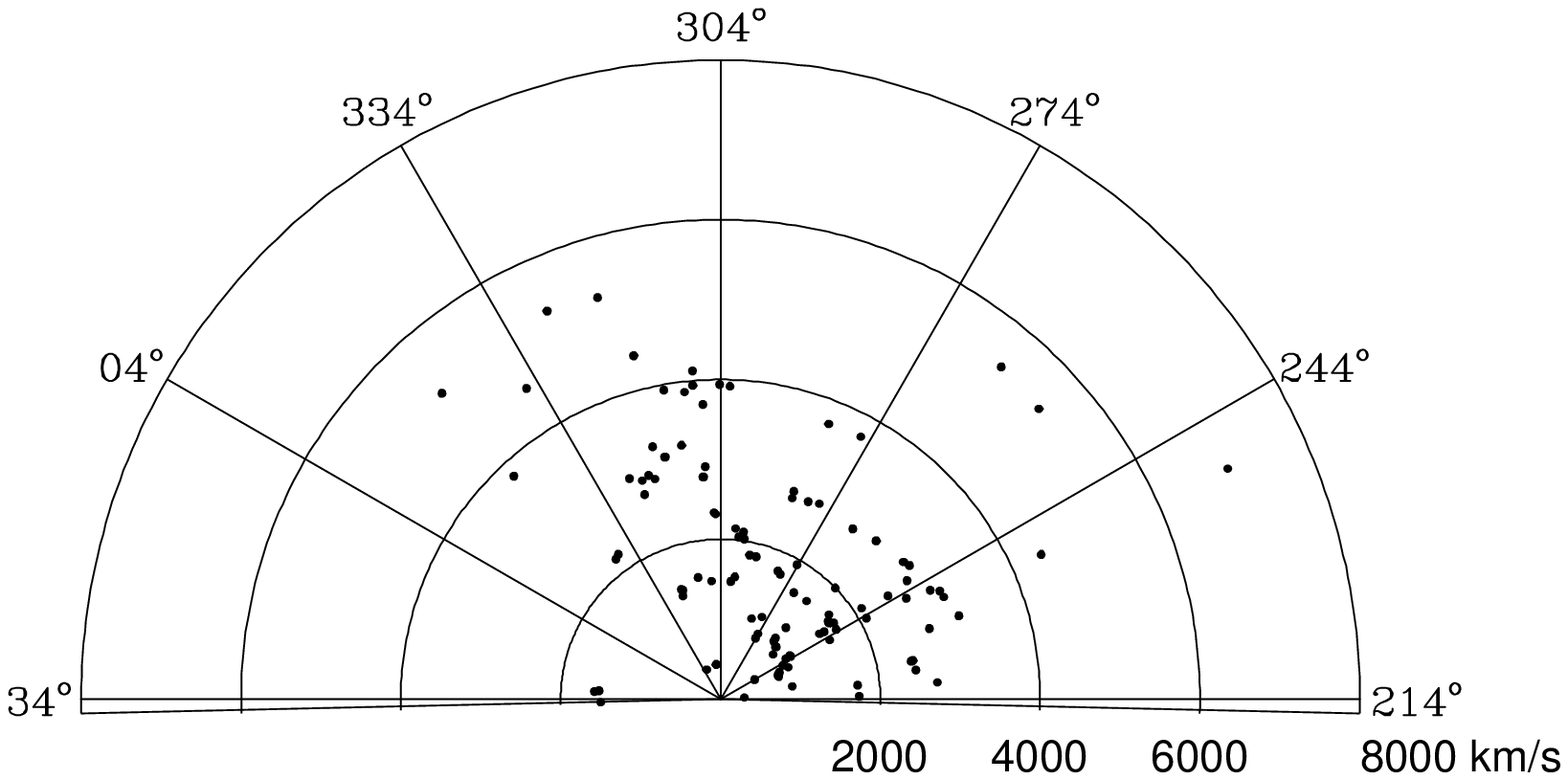,width=17cm}
  \caption{Distribution in Galactic longitude and redshift of the 110 shallow
survey detections.}
  \label{fig-4}
  \end{figure}
The three shallow survey objects on the border of the Local Void
at l $\sim$ 30$^\circ$ all have redshifts $\sim$ 1500 \kms, consistent
with their being members of the group proposed by Roman \etal (1998).
Two of these galaxies lie within the survey boundary of the Dwingeloo
Obscured Galaxies Survey, and were also detected by this survey
(Rivers, Henning, \& Kraan-Korteweg 1999).

The most obvious overdensity of the low latitude galaxies apparent
in Figure 2 occurs at
l $\sim$ 245$^\circ$.
Figure 4 reveals
the overdensity is caused by two groupings of galaxies, at
(l, v) $\sim$ (245$^\circ$, 800 \kms) and
(l, v) $\sim$ (245$^\circ$, 1500 \kms).
The latter corresponds to
the location of the moderately obscured, nearby cluster in 
Puppis (l, b, v) $\sim$ (245$^\circ$, 0$^\circ$, 1500 \kms)
(Kraan-Korteweg \& Huchtmeier 1992).
This overdensity of galaxies was deduced by Scharf \etal (1992)
through a spherical harmonic analysis of IRAS galaxies.
The nearer group was also recovered by Kraan-Korteweg \& Huchtmeier (1992).
There is no evidence of any other nearby, unrecognized clusters of
galaxies in the survey region.
Except near the Galactic center where continuum emission decreased
the sensitivity of the survey somewhat,
the survey was uniformly sensitive to spirals with 
M$_{\rm HI}$ = $2 \times 10^9$
M$_{\odot}$ at 1500 \kms (H$_0$ = 75 \kms Mpc$^{-1}$) across the entire
southern ZOA, so any hidden overdensities within this redshift range
should have been easily detected.

Structures at v $\sim$ 4000 \kms are not well probed by this
survey, however we can see the relative overdensity of galaxies
toward the general Great Attractor region, and galaxy underdensity
behind the Puppis cluster, broadly consistent with the theoretical mass
density reconstructions of Kolatt, Dekel, \& Lahav (1995) and Webster,
Lahav, \& Fisher (1997).
The deep survey should definitively confirm or refute the predictions
of these and other mass density reconstructions in the ZOA.

\subsection{\HI Mass Function}

To determine the number density of galaxies as a function
of \HI mass (the \HI mass function), the sensitivity of the
survey must be carefully determined to be able to apply 
appropriate volume corrections.
Near strong continuum sources, galaxies which would
otherwise be detected are hidden in the increased noise.
Over the search area,
$212^\circ \le \ell \le 36^\circ$, $|b|$~$\leq$~5$^\circ$,
7.7\% of the data were disturbed by strong continuum emission, and
had rms noise fluctuations a factor of three or more above the
quoted sensitivity of 15 mJy.
In the volume correction calculations, we take the survey
area to be 92.3\% of the area covered by the telescope.

The galaxy selection function for the shallow survey is not
based simply on peak flux, but is determined empirically
to involve both total flux and linewidth in the following 
way:  $\int$ S dv $\times$ W$_{50}^{-0.7}$ $>$ 0.3.
Schneider, Spitzak, \& Rosenberg (1998) find a very
similar functional form for the completeness limit of
two previous blind 21-cm surveys done at Arecibo, 
$\int$ S dv $\times$ W$_{50}^{-0.75}$.
With this description, the maximum volume in which each galaxy
could be detected is calculated.
The average value of V / V$_{max}$ = 0.58, is quite close to the
theoretical value of 1/2 for a correctly determined completeness.
The resulting number density of \HI masses with
each galaxy weighted by the inverse of its maximum detectable 
volume is shown in Figure 5.
   \begin{figure}
   \psfig{file=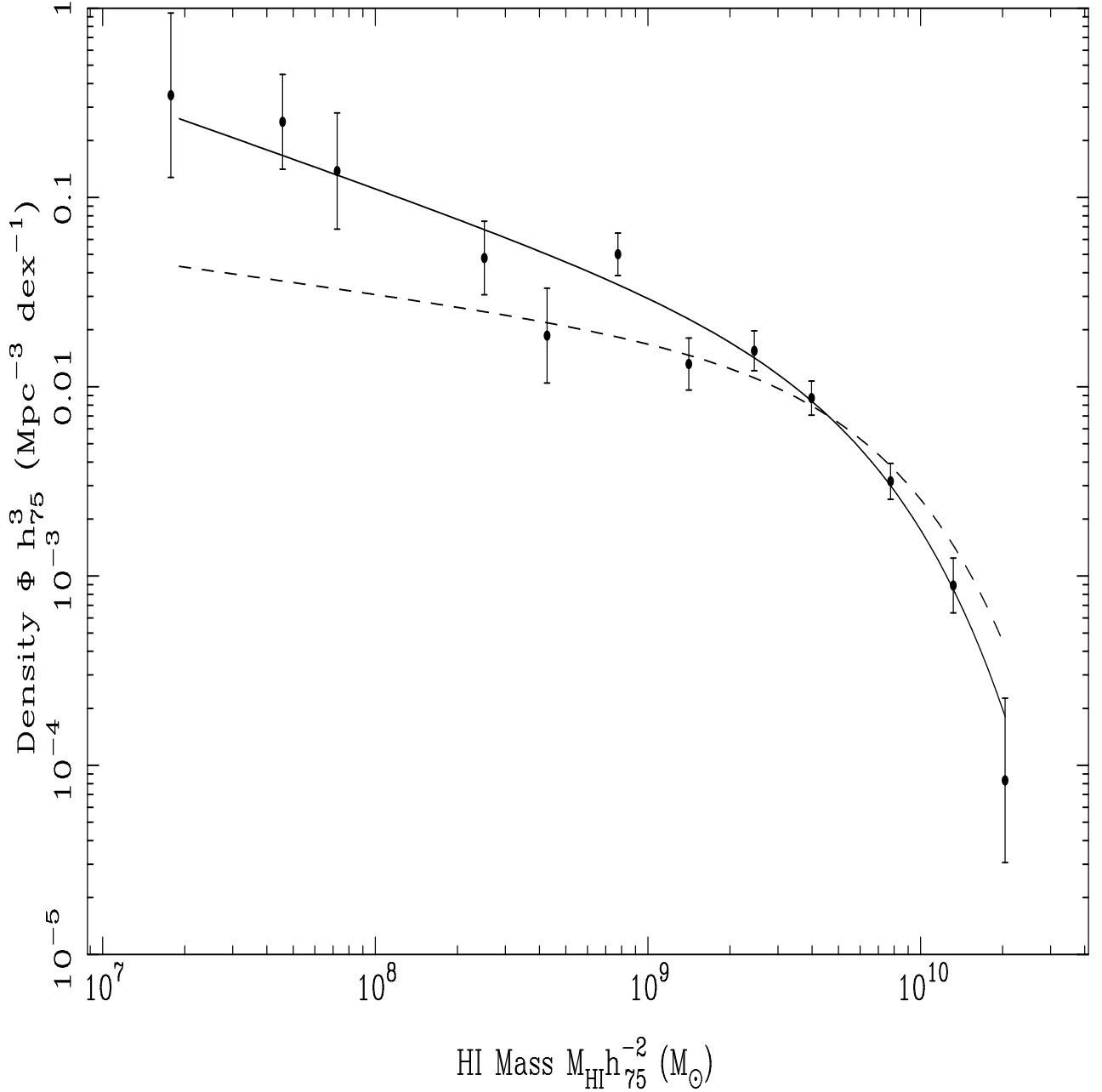,width=17cm}
\caption{\HI mass function determined from the shallow survey
detections.
The errorbars on the points are determined by counting statistics.
The function is satisfactorily fit by a Schechter function
with $\alpha$ = 1.51 $\pm$ 0.12, $\Phi^*$ = 0.006 $\pm$ 0.003,
and log M$^*$ = 9.7 $\pm$ 0.10, shown by the solid curve.  The dashed
curve shows an \HI mass function derived by Zwaan \etal (1997) for
a smaller \HI -selected sample of galaxies.}
   \label{fig-5}
   \end{figure}
This \HI mass function is well fit by a Schechter function (Schechter 1976)
with parameters $\alpha$ = 1.51 $\pm$ 0.12, $\Phi^*$ = 0.006 $\pm$ 0.003, 
and log M$^*$ = 9.7 $\pm$ 0.10.
Note that the Schechter function was not used as an {\it a priori}
assumption of the shape of the \HI mass function.
The low-mass slope is significantly steeper than the $\alpha$ = 1.20 
suggested by Zwaan \etal (1997).
The densities at the lowest mass bins are consistent with the higher
densities found by Schneider, Spitzak, \& Rosenberg (1998).
The main difficulty remains small number statistics, with the
three lowest mass bins of Figure 5 containing a total of six galaxies. 
With the increased search volume of the HIPASS and full sensitivity
ZOA surveys, the statistical robustness 
of the mass function parametrization will be improved.

\subsection{Predictions for the Full Sensitivity Survey}

We now estimate how many galaxies should be uncovered
by the full sensitivity ZOA survey, to be completed in the year 2000.
The deep survey will consist of 435 scans at each longitude,
compared with the 34 of the shallow survey.
This factor of 12.5 increase in integration time will
lead to a $\sqrt{12.5}$ = 3.5 improvement in sensitivity.
For a galaxy of a given M$_{\rm HI}$ and velocity linewidth,
the distance to which it could be detected will increase by
a factor of $\sqrt{3.5}$.
Thus, the volume increase of the deep survey over the
shallow survey is about a factor of 6.5,
leading to a rough estimate of $6.5 \times 110 \approx 700$ galaxies
to be detected by the deep survey.
Indeed, a portion of the deep survey is completed, and the four
data cubes in the region of the supposed Great Attractor have
been inspected, and about 300 galaxy candidates noted
(S. Juraszek, private communication).
Extrapolation over the full spatial extent of the survey
leads to an estimate of about 1700 galaxies.
However, this region of space contains a significant
overdensity of galaxies, and much of the final survey volume will
contain the Local Void, so the total tally may be closer to
1000 galaxies.

Efforts are underway to develop software tools which
model and remove strong continuum sources from the data, which
should decrease the effective noise even further, increasing the
survey sensitivity and the number of detected objects.

\section*{Acknowledgements}

We thank the HIPASS team (PI:  Rachel Webster)
for assisting with the observing,
and the staff at Parkes for their support.
We also acknowledge the {\sc aips++} group
for the development of the basis of the data reduction software and
some of the observing software.
David Barnes contributed very useful software for the analysis of cube
statistics.
We are grateful to the multibeam instrument teams, headed by Warwick
Wilson, Mal Sinclair, and Trevor Bird.

This research has made use of the NASA/IPAC Extragalactic Database (NED)
which is operated by the Jet Propulsion Laboratory, Caltech, under
contract with the National Aeronautics and Space Administration.
We have also made use of the Lyon-Meudon Extragalactic Database (LEDA),
supplied by the LEDA team at the Centre de Recherche Astronomique de
Lyon, Observatoire de Lyon.
The research of P.H. is supported by NSF Faculty Early Career Development
(CAREER) Program award AST 95-02268.
P.H. warmly thanks the ATNF for the hospitality and support during her 
sabbatical stay.


\begin{references}


\reference{} Barnes, D.G. 1998, in ASP Conf. Ser. 145, Astronomical Data
Analysis Software and Systems VII, ed. R. Albrecht, R.N. Hook, \&
H.A. Bushouse (San Francisco:  ASP), 32

\reference{} Barnes, D.G., Staveley-Smith, L.,  Ye, T., \& Oosterloo, T. 1998,
in ASP Conf. Ser. 145, Astronomical Data Analysis Software and Systems VII,
ed. R. Albrecht, R.N. Hook, \& H.A. Bushouse (San Francisco:  ASP), 32

\reference {} Barnes, D.G., \etal 2000, in preparation

\reference{} Epchtein, N. 1997, in The Impact of Large Scale Near-Infrared
Surveys, ed. F. Garzon \etal (Dordrecht:  Kluwer), 15

\reference{} Garcia, A.M, Bottinelli, L., Garnier, R., Gouguenheim, L.,
\& Paturel, G. 1994, A\&AS, 107, 265

\reference{} Gooch, R.E. 1995, in ASP Conf. Ser. 101, 
Astronomical Data Analysis Software and Systems V, ed. G.H. Jacoby \&
J. Barnes (San Francisco:  ASP), 80

\reference{} Henning, P.A., Kraan-Korteweg, R.C., Rivers, A.J., Loan,
A.J., Lahav, O., \& Burton, W.B. 1998, AJ, 115, 584

\reference{} Huchtmeier, W.K., \& Richter, O.G. 1986, 
A\&AS, 63, 323
 
\reference{} Juraszek, S., Staveley-Smith, L., Kraan-Korteweg, R.C., Green, A.J.,
Ekers, R.D., Haynes, R.F., Henning, P., Kesteven, M.J., Koribalski, B., 
Price, R.M., Sadler, E.M., 
\& Schr\"{o}der, A. 2000, AJ, in press

\reference{} Kerr, F.J., \& Henning, P.A. 1987, ApJ, 320, L99

\reference{} Kolatt, T., Dekel, A., \& Lahav, O. 1995, MNRAS, 275, 797

\reference{} Kraan-Korteweg, R.C., \& Huchtmeier, W.K. 1992, A\&A,
266, 150

\reference{} Kraan-Korteweg, R.C. 1993, in The Distribution of Matter in
the Universe, ed. G. Mamon \& D. Gerbal, 
(Paris:  Observatoire de Paris Press), 202

\reference{} Kraan-Korteweg, R.C., Loan, A.J., Burton, W.B., Lahav, O.,
Ferguson, H.C., Henning, P.A., \& Lynden-Bell, D. 1994, Nature, 372, 77

\reference{} Kraan-Korteweg, R.C., \& Woudt, P.A. 1999, PASA, 16, 53

\reference{} Kraan-Korteweg, R.C., \& Juraszek, S. 2000, PASA, in press

\reference{} Lauberts, A. 1982, The ESO/Uppsala Survey of the ESO (B) Atlas,
(Garching:  ESO)

\reference{} Lewis, B.M. 1983, AJ, 88, 962 

\reference{} Rivers, A.J., Henning, P.A., \& Kraan-Korteweg, R.C. 1999,
PASA, 16, 48

\reference{} Roman, A.T., Takeuchi, T.T., Nakanishi, K., \& Sait\={o}, M.
1998, Publ. Astron. Soc. of Japan, 50, 47

\reference{} Sault, R.J., Teuben, P.J., \& Wright, M.C.H. 1995, in
ASP Conf. Ser. 77, Astronomical Data Analysis Software and Systems IV,
ed. R. Shaw, H.E. Payne, J.J.E. Hayes (San Francisco: ASP), 433


\reference{} Scharf, C., Hoffman, Y., Lahav, O., \& Lynden-Bell, D.
1992, MNRAS, 256, 229

\reference{} Schechter, P. 1976, ApJ, 203, 297

\reference{} Schlegel, D.J., Finkbeiner, D.P. \& Davis, M. 1998 ApJ, 500, 525

\reference{} Schneider, S.E., Spitzak, J.G., \& Rosenberg, J.L. 1998,
ApJ, 507, L9

\reference{} Schr\"{o}der, A., Kraan-Korteweg, R.C., \& Mamon, G.A. 1999,
PASA, 16, 42

\reference{} Skrutskie, M.F., \etal 1997, in The Impact of Large Scale Near-Infrared
Surveys, ed. F. Garzon \etal (Dordrecht:  Kluwer), 25 

\reference{} Staveley-Smith, L., \& Davies, R.D. 1987, MNRAS, 224, 953

\reference{} Staveley-Smith, L., \& Davies, R.D. 1988, MNRAS, 231, 833

\reference{} Staveley-Smith, L. 1997, PASA,
14, 111

\reference{} Staveley-Smith, L., Juraszek, S., Koribalski, B.S.,  Ekers, R.D.,
Green, A.J., Haynes, R.F.,  Henning, P.A., Kesteven, M.J.,  Kraan-Korteweg,
R.C., Price, R.M. \& Sadler, E.M. 1998, AJ, 116, 2717

\reference{} Takata, T., Yamada, T., Sait\={o}, M., Chamaraux, P., \& Kaz\'{e}s, I.
1994, A\&AS, 104, 529

\reference{} Theureau, G., Bottinelli, L., Coudreau-Durand, N., Gouguenheim, L.,
Hallet, N., Loulergue, M., Paturel, G., \& Teerikorpi, P. 1998, A\&AS, 130, 333

\reference{} Webster, M., Lahav, O., \& Fisher, K. 1997, MNRAS, 287, 425

\reference{} Zwaan, M.A., Briggs, F.H., Sprayberry, D., \& Sorar, E.
1997, ApJ, 490, 173
 
\end{references}
\end{document}